\documentclass[twocolumn,showpacs,aps,prc,floatfix,superscriptaddress]{revtex4}
\usepackage{bm}
\usepackage{graphicx}

\usepackage{color}
\begin{document}

\title{
Spin alignment measurements of the $K^{*0}(892)$ and $\phi(1020)$
vector mesons in heavy ion collisions at $\sqrt{s_{NN}}$=200~GeV}

\affiliation{Argonne National Laboratory, Argonne, Illinois 60439}
\affiliation{University of Birmingham, Birmingham, United Kingdom}
\affiliation{Brookhaven National Laboratory, Upton, New York
11973} \affiliation{California Institute of Technology, Pasadena,
California 91125} \affiliation{University of California, Berkeley,
California 94720} \affiliation{University of California, Davis,
California 95616} \affiliation{University of California, Los
Angeles, California 90095} \affiliation{Universidade Estadual de
Campinas, Sao Paulo, Brazil} \affiliation{Carnegie Mellon
University, Pittsburgh, Pennsylvania 15213}
\affiliation{University of Illinois at Chicago, Chicago, Illinois
60607} \affiliation{Creighton University, Omaha, Nebraska 68178}
\affiliation{Nuclear Physics Institute AS CR, 250 68
\v{R}e\v{z}/Prague, Czech Republic} \affiliation{Laboratory for
High Energy (JINR), Dubna, Russia} \affiliation{Particle Physics
Laboratory (JINR), Dubna, Russia} \affiliation{University of
Frankfurt, Frankfurt, Germany} \affiliation{Institute of Physics,
Bhubaneswar 751005, India} \affiliation{Indian Institute of
Technology, Mumbai, India} \affiliation{Indiana University,
Bloomington, Indiana 47408} \affiliation{Institut de Recherches
Subatomiques, Strasbourg, France} \affiliation{University of
Jammu, Jammu 180001, India} \affiliation{Kent State University,
Kent, Ohio 44242} \affiliation{University of Kentucky, Lexington,
Kentucky, 40506-0055} \affiliation{Institute of Modern Physics,
Lanzhou, China} \affiliation{Lawrence Berkeley National
Laboratory, Berkeley, California 94720} \affiliation{Massachusetts
Institute of Technology, Cambridge, MA 02139-4307}
\affiliation{Max-Planck-Institut f\"ur Physik, Munich, Germany}
\affiliation{Michigan State University, East Lansing, Michigan
48824} \affiliation{Moscow Engineering Physics Institute, Moscow
Russia} \affiliation{City College of New York, New York City, New
York 10031} \affiliation{NIKHEF and Utrecht University, Amsterdam,
The Netherlands} \affiliation{Ohio State University, Columbus,
Ohio 43210} \affiliation{Panjab University, Chandigarh 160014,
India} \affiliation{Pennsylvania State University, University
Park, Pennsylvania 16802} \affiliation{Institute of High Energy
Physics, Protvino, Russia} \affiliation{Purdue University, West
Lafayette, Indiana 47907} \affiliation{Pusan National University,
Pusan, Republic of Korea} \affiliation{University of Rajasthan,
Jaipur 302004, India} \affiliation{Rice University, Houston, Texas
77251} \affiliation{Universidade de Sao Paulo, Sao Paulo, Brazil}
\affiliation{University of Science \& Technology of China, Hefei
230026, China} \affiliation{Shanghai Institute of Applied Physics,
Shanghai 201800, China} \affiliation{SUBATECH, Nantes, France}
\affiliation{Texas A\&M University, College Station, Texas 77843}
\affiliation{University of Texas, Austin, Texas 78712}
\affiliation{Tsinghua University, Beijing 100084, China}
\affiliation{Valparaiso University, Valparaiso, Indiana 46383}
\affiliation{Variable Energy Cyclotron Centre, Kolkata 700064,
India} \affiliation{Warsaw University of Technology, Warsaw,
Poland} \affiliation{University of Washington, Seattle, Washington
98195} \affiliation{Wayne State University, Detroit, Michigan
48201} \affiliation{Institute of Particle Physics, CCNU (HZNU),
Wuhan 430079, China} \affiliation{Yale University, New Haven,
Connecticut 06520} \affiliation{University of Zagreb, Zagreb,
HR-10002, Croatia}

\author{B.I.~Abelev}\affiliation{University of Illinois at Chicago, Chicago, Illinois 60607}
\author{M.M.~Aggarwal}\affiliation{Panjab University, Chandigarh 160014, India}
\author{Z.~Ahammed}\affiliation{Variable Energy Cyclotron Centre, Kolkata 700064, India}
\author{B.D.~Anderson}\affiliation{Kent State University, Kent, Ohio 44242}
\author{D.~Arkhipkin}\affiliation{Particle Physics Laboratory (JINR), Dubna, Russia}
\author{G.S.~Averichev}\affiliation{Laboratory for High Energy (JINR), Dubna, Russia}
\author{Y.~Bai}\affiliation{NIKHEF and Utrecht University, Amsterdam, The Netherlands}
\author{J.~Balewski}\affiliation{Massachusetts Institute of Technology, Cambridge, MA 02139-4307}
\author{O.~Barannikova}\affiliation{University of Illinois at Chicago, Chicago, Illinois 60607}
\author{L.S.~Barnby}\affiliation{University of Birmingham, Birmingham, United Kingdom}
% ...

\collaboration{STAR Collaboration}\noaffiliation


\begin{abstract}

We present the first spin alignment measurements for the
$K^{*0}(892)$ and $\phi(1020)$ vector mesons produced at
mid-rapidity with transverse momenta up to 5\,GeV/c at
$\sqrt{s_{NN}}$ = 200\,GeV at RHIC. The diagonal spin density
matrix elements with respect to the reaction plane in Au+Au
collisions are $\rho_{00}$ = 0.32 $\pm$ 0.04 (stat) $\pm$ 0.09
(syst) for the $K^{*0}$ ($0.8<p_T<5.0$~GeV/c) and $\rho_{00}$ =
0.34 $\pm$ 0.02 (stat) $\pm$ 0.03 (syst) for the $\phi$
($0.4<p_T<5.0$ GeV/c), and are constant with transverse momentum
and collision centrality. The data are consistent with the
unpolarized expectation of 1/3 and thus no evidence is found for
the transfer of the orbital angular momentum of the colliding
system to the vector meson spins. Spin alignments for $K^{*0}$ and
$\phi$ in Au+Au collisions were also measured with respect to the
particle's production plane. The $\phi$ result, $\rho_{00}$ = 0.41
$\pm$ 0.02 (stat) $\pm$ 0.04 (syst), is consistent with that in
p+p collisions, $\rho_{00}$ = 0.39 $\pm$ 0.03 (stat) $\pm$ 0.06
(syst), also measured in this work. The measurements thus
constrain the possible size of polarization phenomena in the
production dynamics of vector mesons.

\end{abstract}
\maketitle

Measurements of the polarization of the particles produced in
relativistic heavy-ion collisions may provide new insights into
the initial conditions and evolution of the nuclear system during
the collision~\cite{LiangPRL,Polarization-New,Polarization-New2}.
In particular, by studying the polarization magnitudes with
respect to various kinematic planes one could attempt to discern
the point in the evolution of the system at which the polarization
arises and, hence, the dominant mechanisms involved. The planes
that are relevant to this paper are the reaction plane, which is
defined by the beam momentum and the nuclear impact parameter, and
the particle's production plane, which is defined by the beam
momentum and the momentum of the produced particle.

In non-central relativistic heavy-ion collisions, transverse
gradients of the total longitudinal momentum of the participant
matter result in substantial local orbital angular momentum of the
created partons~\cite{LiangPRL}. Due to the spin-orbit coupling of
QCD, this orbital motion may result in a net polarization of the
produced particles along the direction of the initial angular
momentum, that is, perpendicular to the reaction plane, yielding a
global hadronic polarization in the final
state~\cite{LiangPRL,LiangQM06,Sergei}. The magnitude and the
transverse momentum ($p_{T}$) dependence of the global
polarization are therefore expected to be sensitive to different
hadronization scenarios~\cite{LiangPLB}. In particular, the
proposed quark recombination model for hadronization of bulk
partonic matter created at RHIC~\cite{Recom-1}, which reproduces
measurements in the intermediate $p_{T}$ region
($2<p_{T}<5$~GeV/c) quite well~\cite{STAR-white-paper}, may be an
effective dynamical mechanism for transferring polarization from
quarks to vector mesons or hyperons.

One can study the polarization of final state hadrons with respect
to the particle's production plane as well. Non-zero polarizations
transverse to this plane are expected to be sensitive to particle
formation dynamics and to possible intrinsic quark transverse spin
distributions~\cite{Hyperons-polarizations-list-theory}. Because
large production plane polarizations have been observed for
hyperons in unpolarized p+p and p+A
interactions~\cite{Hyperons-polarizations-list-data} and for
vector mesons in K+p, n+C, and $e^+e^-$
interactions~\cite{vector-meson-1,vector-meson-2,LEP-DELPHI,LEP-OPAL},
the disappearance of these effects in relativistic heavy-ion
collisions might indicate that the system is isotropic to the
extent that, locally, there is no longer a preferred
direction~\cite{QGP-Unpolarized}. It has also been
suggested~\cite{Qinhua-vector-meson} that vector meson spin
alignment with respect to the production plane is closely related
to the single-spin left-right asymmetries in transversely
polarized p+p
collisions~\cite{Fermilab-AN,STAR-ANMR,STAR-AN,PHENIX-AN}.

We have recently measured the global (reaction plane) polarizations
of $\Lambda$ and $\bar{\Lambda}$ hyperons produced at mid-rapidity
in Au+Au collisions at center-of-mass energies $\sqrt{s_{NN}} =
62.4\,\mathrm{GeV}$ and 200\,GeV~\cite{STAR-Lambda}. The results,
$\left| P_{\Lambda,\bar{\Lambda}} \right| \leq 0.02$, are well below
the predictions of Ref.~\cite{LiangPRL} and are in agreement with
the refined calculations of Ref.~\cite{LiangQM06}.

In this paper, we present the first measurements of the spin
alignment for the $K^{*0}$ and $\phi$ vector mesons with respect
to both the reaction plane and the production plane in  Au+Au
collisions at $\sqrt{s_{NN}}$ = 200\,GeV. We present also the spin
alignment of the $\phi$ meson with respect to its production plane
for p+p collisions at the same collision energy. The $K^{*0}$ data
cover $0.8<p_T<5.0$~GeV/c and the $\phi$ data cover
$0.4<p_T<5.0$~GeV/c.

Spin alignment is described by a spin density matrix $\rho$, a
$3\times3$ Hermitian matrix with unit trace. A deviation of the
diagonal elements  $\rho_{mm}(m=-1,0,1)$ from 1/3 signals net spin
alignment. Because vector mesons decay strongly, the diagonal
elements $\rho_{-1-1}$ and $\rho_{11}$ are degenerate and
$\rho_{00}$ is the only independent observable. It can be
determined from the angular distribution of the decay
products~\cite{eqa-theta},
\begin{equation}
\frac{dN}{d(\cos\theta^{*})}=N_0 \times
[(1-\rho_{00})+(3\rho_{00}-1)\cos^{2}\theta^{*}], \label{eq-theta}
\end{equation}
where $N_0$ is the normalization and $\theta^{*}$ is the angle
between the polarization direction $\hat n$ and the momentum
direction of a daughter particle in the rest frame of the parent
vector meson. In the case of a global spin alignment measurement,
the polarization direction $\hat n$ is along the orbital angular
momentum of the colliding system. It is determined by the reaction
plane, requiring only the second order term since Eq.(1) is
invariant under $\theta^{*} \rightarrow \pi +
\theta^{*}$~\cite{Art}. For the production plane measurement,
$\hat n$ lies along the normal to the production plane, which is
determined by the momentum of the vector meson and of the
colliding beams.

Vector mesons are expected to originate predominantly from
primordial
production~\cite{Primodial-production-1,Primodial-production-2},
unlike hyperon production which is expected to have large
resonance decay
contributions~\cite{STAR-Lambda,Primodial-production-1,Primodial-production-2}.
Another difference between the present spin alignment measurement
and our recent measurement of global hyperon
polarization~\cite{STAR-Lambda} is that contributions to a spin
alignment measurement are generally additive, whereas
contributions along a polarization direction may cancel. Last, as
far as the reaction plane resolution is concerned, the
aforementioned method has an advantage over the method used in
Ref.~\cite{STAR-Lambda}, where the reaction plane was estimated in
forward detectors.


A total of approximately 2.3$\times10^{7}$ events from Au+Au data
collection in the year 2004 run and 6.0$\times 10^{6}$ events from
p+p data collection in the year 2001 run have been used in these
analyses. The events were collected with minimum bias
triggers~\cite{STAR-kstar, STAR-phi}. Charged tracks were
reconstructed with the STAR Time Projection Chamber (TPC) for
pseudo-rapidities $|\eta|<1.0$ and all azimuthal
angles~\cite{STAR-TPC}. Particle identification is achieved by
correlating the ionization energy loss $(dE/dx)$ of charged
particles in the TPC gas with their measured momenta. The measured
$\langle dE/dx \rangle$ is reasonably well described by the Bichsel
function smeared with a resolution of width
$\sigma$~\cite{dEdx-Bichsel}. By measuring the $\langle dE/dx
\rangle$, pions and kaons could be identified up to a momentum of
about 0.6~GeV/c while protons could be separated from pions and
kaons up to a momentum of about 1.1~GeV/c. Tracks within 2$\sigma$
of the pion/kaon Bichsel curve were selected in the analyses. The
$K^{*0}$ and $\phi$ mesons were reconstructed through their
respective hadronic decay channels, $K^{*0}\rightarrow
K^{+}\pi^{-}$, $\overline{K^{*0}}\rightarrow K^{-}\pi^{+}$ and
$\phi\rightarrow K^{+}K^{-}$. The $K^{*0}$ and $\overline{K^{*0}}$
samples were combined to enhance the statistics and the term
$K^{*0}$ in the remainder of this paper will refer to the combined
sample. The collision centrality was determined by the charged
hadron multiplicity within $|\eta|<0.5$. The same analysis
techniques have been used in our earlier measurements of $K^{*0}$
and $\phi$ production~\cite{STAR-kstar, STAR-phi}.

\begin{figure}[htbp]
\includegraphics[scale=0.70,bb=5 10 450 720]{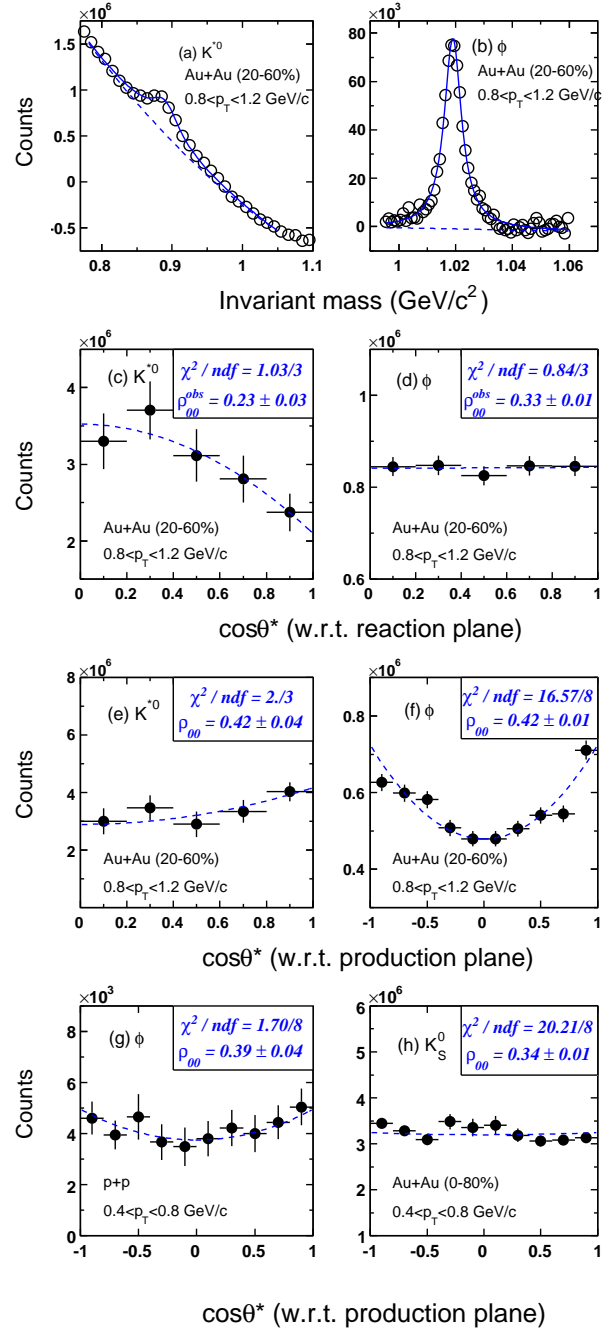}
\caption{\label{fig:InvMass}(color online) The invariant mass
distribution after combinatorial background subtraction for (a)
the $K^{*0}$ and (b) the $\phi$ meson in mid-central Au+Au
collisions at $\sqrt{s_{NN}}$= 200 GeV including all values of
cos$\theta^{*}$. The continuous lines represent fits of signal,
described with a Breit-Wigner function, and residual background,
described with the dashed polynomial curves. Panel (c), (e) and
panel (d), (f) represent the $\cos$$\theta^{*}$ distributions for
the $K^{*0}$ and $\phi$ yields in Au+Au collisions, respectively.
Panel (g) is the $\phi$ yield in p+p collisions while panel (h)
shows the control measurement of the spin-less $K^{0}_{S}$ meson
$\cos$$\theta^{*}$ distribution. The error bars show statistical
uncertainties. The blue dashed lines in (c)-(h) are fits of Eq.
(1) to the data.}
\end{figure}

Figure~\ref{fig:InvMass} illustrates aspects of the data and
analysis for particular $p_{T}$ bins. The top panels show the
invariant mass distributions for $(a)$ $K^{*0}$ and $(b)$ $\phi$
candidates in mid-central (20-60\%) Au+Au collisions at
$\sqrt{s_{NN}}$ = 200\,GeV including all values of
cos$\theta^{*}$. In these analyses invariant mass distributions
were obtained for $K^{*0}$ and $\phi$ for each cos$\theta^{*}$ and
$p_T$ interval. The raw $K^{*0}$ and $\phi$ yields in each of
these distributions were obtained by subtracting the corresponding
combinatorial backgrounds and fitting the remaining distributions
with a Breit-Wigner function plus a polynomial curve to describe
the residual background. The raw yields were then corrected for
detection efficiency and acceptance determined from Monte Carlo
GEANT simulations~\cite{STAR-kstar,STAR-phi}. The middle panels in
Fig.~\ref{fig:InvMass} show the cos$\theta^{*}$ distributions,
after efficiency and acceptance corrections, for the $(c)$
$K^{*0}$ and $(d)$ $\phi$, respectively. Eq.~(\ref{eq-theta}) was
fitted to these distributions to determine $\rho_{00}(p_{T})$. In
the analyses, we used charged particle tracks with
$0.2<p_{T}<2.0$~GeV/c and pseudo-rapidity $|\eta|<1.0$ originating
from the primary interaction vertex to reconstruct the event plane
as an estimate of the reaction plane~\cite{STAR-v2}. Tracks
associated with a $K^{*0}$ or a $\phi$ candidate are explicitly
excluded from the event plane calculation. The results for
$\rho_{00}(p_T)$ were corrected for the finite event plane
resolution, which was determined by correlating two random
sub-events. The correction factor on ($3\rho_{00}-1$) is
determined to be 1/0.81 in 20-60\% Au+Au collisions at
$\sqrt{s_{NN}}$ = 200\,GeV~\cite{STAR-v2}. The bottom panel (e),
(f), (g) represents the cos$\theta^{*}$ distribution for $K^{*0}$
and $\phi$ mesons with respect to the production plane in Au+Au
and p+p collisions. In this case, $\rho_{00}(p_T)$ is extracted
directly by fitting Eq.~(\ref{eq-theta}) to the distributions. We
have checked our analysis procedure by extracting $\rho_{00}$ for
the abundantly produced, but spin-less $K^{0}_{S}$ meson
$(J^P=0^-)$. The results are shown in panel (h) and are consistent
with 1/3 within the statistical uncertainties, as expected. The
$\chi^2/ndf$ value is unsatisfactory for the $K^{0}_{S}$ fit for
$0.4<p_T<0.8$~GeV/c, which is indicative of point-to-point
systematics. It reaches satisfactory values at larger $p_T$.

\begin{figure}[htb!]
\includegraphics[scale=0.45,bb=20 10 850 400]{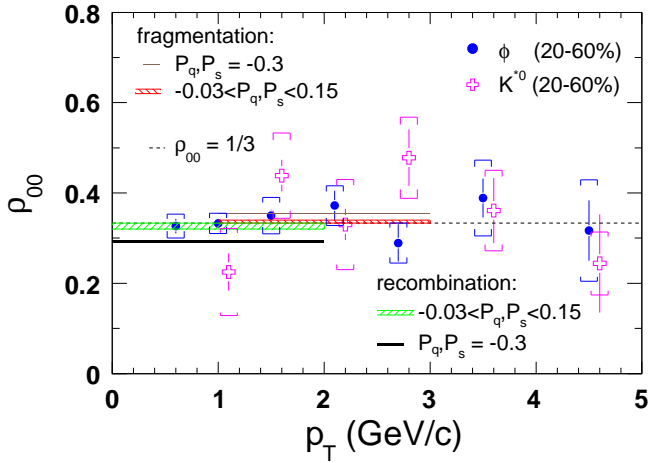}
\vspace{-0.55cm} \caption{\label{fig:rho00-RP}(color online) The
spin density matrix elements $\rho_{00}$ with respect to the
reaction plane in mid-central Au+Au collisions at $\sqrt{s_{NN}}$
= 200 GeV versus $p_T$ of the vector meson. The sizes of the
statistical uncertainties are indicated by error bars, and the
systematic uncertainties by caps . The $K^{*0}$ data points have
been shifted slightly in $p_T$ for clarity. The dashed horizontal
line indicates the unpolarized expectation $\rho_{00} = 1/3$. The
bands and continuous horizontal lines show predictions discussed
in the text. }
\end{figure}

\begin{table}
\centering \caption{The averaged spin density matrix elements
$\rho_{00}$ with respect to the reaction plane in mid-central
Au+Au collisions at $\sqrt{s_{NN}}$ = 200 GeV.} \label{table1}
\begin{ruledtabular}
\begin{tabular}{c| c | c}
            &     $K^{*0}$    &    $\phi$
\\ \hline
$\rho_{00}(p_{T}<2.0$~GeV/c) & 0.31 $\pm$ 0.04 $\pm$ 0.09 & 0.33
$\pm$ 0.01 $\pm$ 0.03
\\ \hline
$\rho_{00}(p_{T}>2.0$~GeV/c) & 0.37 $\pm$ 0.04 $\pm$ 0.09 & 0.35
$\pm$ 0.04 $\pm$ 0.05
\\ \hline
$\rho_{00}(p_{T}<5.0$~GeV/c) & 0.32 $\pm$ 0.03 $\pm$ 0.09 & 0.34
$\pm$ 0.02 $\pm$ 0.03
\end{tabular}
\end{ruledtabular}
\end{table}

The measurements of the $K^{*0}$ and $\phi$ global spin alignment
versus $p_{T}$ of the vector meson for mid-central Au+Au
collisions at $\sqrt{s_{NN}}$ = 200~GeV are presented in
Fig.~\ref{fig:rho00-RP}, and are summarized in Table~\ref{table1}.
Both statistical and systematic uncertainties are shown. The
dominant contribution to the systematic uncertainty for the $\phi$
($K^{*0}$) meson ranges from 0.020 (0.05) to 0.045 (0.10),
originating from uncertainty in the magnitude and shape of the
residual background after the subtraction of combinatorial
background. This residual arises from the incomplete description
of combinatorial background via the event mixing technique and
from distortions to the background in the invariant mass
distribution near the $\phi$ peak caused by photon conversions and
other correlated backgrounds such as $K^{0*}\rightarrow K^+\pi^-$,
$\rho^0\rightarrow \pi^+\pi^-$, $\Lambda\rightarrow p\pi^-$ and
$\Delta\rightarrow N\pi$ decays~\cite{JGMa-Thesis}. In the case of
the $K^{*0}$ these backgrounds include $K_{S}^{0}\rightarrow
\pi^+\pi^-$, $\rho^0\rightarrow \pi^+\pi^-$, $\phi\rightarrow
K^+K^-$, $\Lambda\rightarrow p\pi^-$ and $\Delta\rightarrow N\pi$
decays~\cite{Haibin-Thesis}. Other point-to-point systematic
uncertainty associated with particle identification for the $\phi$
($K^{*0}$) meson were estimated to range from 0.007 (0.06) to
0.012 (0.09) by tightening the $K^{\pm}$ ($\pi$ and $K$) $\langle
dE/dx \rangle$ cut from 2$\sigma$ to 1$\sigma$. An additional
sizable contribution to the $\phi$ uncertainty was estimated to
range from 0.007 to 0.012 by varying the fitted invariant mass
range from 1.00--1.04~GeV/$c^{2}$ to 1.00--1.06~GeV/$c^{2}$, and
to the $K^{*0}$ uncertainty ranging from 0.02 to 0.05 by changing
its analyzed rapidity range from $|y|<1$ to $|y|<0.5$. The
systematic uncertainties in the $K^{0*}$ measurements are larger
than those in the $\phi$ measurement mainly because of the lower
signal-to-background ratio of $\sim 1/1000$ compared to $\sim1/25$
for the $\phi$ meson. The contributions to the systematic
uncertainty caused by elliptic flow effects and the event plane
resolution are found to be negligible. The $K^{*0}$ and $\phi$
data are consistent with each other and are consistent with 1/3 at
all $p_T$.

Hadronization of globally polarized thermal quarks, typically
having $p_T < 1\,\mathrm{GeV/c}$, in mid-central Au+Au collisions
is predicted to cause $p_T$ dependent deviations of $\rho_{00}$
from the unpolarized value of
1/3~\cite{LiangPRL,LiangQM06,LiangPLB,Liang-Private}.
Recombination of polarized thermal quarks and anti-quarks is
expected to dominate for $p_{T}<2$\,GeV/c and leads to values of
$\rho_{00}$ $<1/3$ as indicated in Fig.~\ref{fig:rho00-RP} for a
typical range of expected light (strange) quark polarizations
$P_{q(s)}$~\cite{LiangPLB}. The fragmentation of polarized thermal
quarks with larger $p_T$, however, would lead to values of
$\rho_{00}$ $>1/3$ for
$1<p_T<3$~GeV/c~\cite{LiangPLB,Liang-Private}, which is indicated
as well. In the region of $1<p_T<2$~GeV/c both hadronization
mechanisms could occur and their effects on $\rho_{00}$ may
cancel. As observed in Fig.~\ref{fig:rho00-RP} these effects are
predicted to be smaller than our experiment sensitivity. However,
the large (strange) quark polarization, $P_{q,s} = -0.3$,
considered in the recombination scenario of Ref.~\cite{LiangPRL}
results in worse agreement of $\rho_{00}$ with our $\phi$ data
than $-0.03<P_{q,s}<0.15$ discussed in Ref.~\cite{LiangQM06}. Our
data are consistent with the unpolarized expectation
$\rho_{00}=1/3$. Recent measurement of the $\Lambda$ and
$\bar\Lambda$ global polarization also found no significant
polarization and an upper limit,
$|P_{\Lambda,\overline{\Lambda}}|\leq0.02$, was
obtained~\cite{STAR-Lambda}.

\begin{figure}[htb!]
\includegraphics[scale=0.45,bb=20 10 850 400]{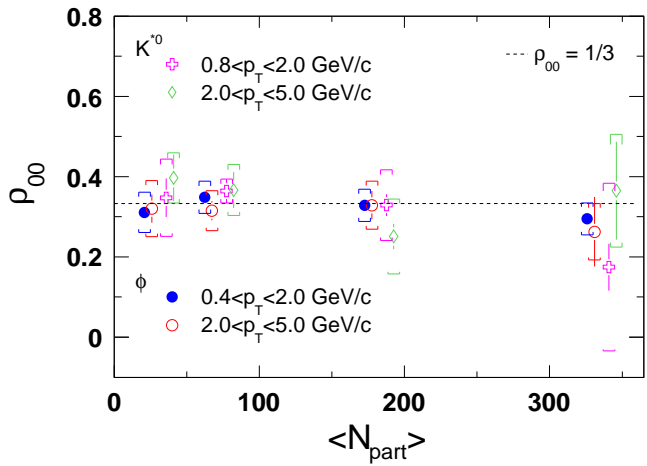}
\vspace{-0.55cm} \caption{\label{fig:rho00-npart-RP}(color online)
The dependence of $\rho_{00}$ with respect to the reaction plane
on the number of participants at mid-rapidity in Au+Au collisions
at $\sqrt{s_{NN}}$ = 200 GeV. The sizes of the statistical
uncertainties are indicated by error bars, and the systematic
uncertainties by caps. The $\phi$ data for $p_{T}>2$~GeV/c and the
$K^{*0}$ data points have been shifted slightly in $\langle
N_{part} \rangle$ for clarity. The dashed horizontal line
indicates the unpolarized expectation $\rho_{00} = 1/3$. }
\end{figure}

The centrality dependence of the global spin alignment
measurements for $K^{*0}$ and $\phi$ vector mesons with low and
intermediate $p_T$ is shown in Fig.~\ref{fig:rho00-npart-RP}. The
orbital angular momentum of the colliding system depends strongly
on the collision centrality. Global polarization is predicted to
be vanishingly small in central collisions and to increase almost
linearly with impact parameter in semi-central collisions due to
increasing particle angular momentum along with effects of
spin-orbit coupling in QCD~\cite{LiangPRL}. The data exhibit no
significant spin alignment at any collision centrality and thus
can constrain the possible size of spin-orbit couplings.

\begin{figure}
\includegraphics[scale=0.45,bb=20 10 850 400]{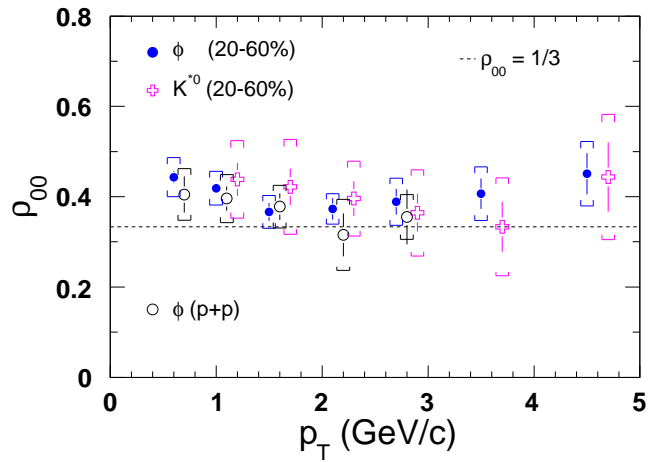}
\vspace{-0.55cm} \caption{\label{fig:rho00-PP}(color online) The
spin density matrix elements $\rho_{00}$ with respect to the
production plane in mid-central Au+Au and p+p collisions at
$\sqrt{s_{NN}}$ = 200~GeV versus $p_{T}$ of the vector meson. The
sizes of the statistical uncertainties are indicated by error
bars, and the systematic uncertainties by caps. The $K^{*0}$ and
the $\phi$ p+p data points have been shifted slightly in $p_T$ for
clarity. The dashed horizontal line indicates the unpolarized
expectation $\rho_{00} = 1/3$. }
\end{figure}

Figure~\ref{fig:rho00-PP} and Table~\ref{table2} present the
$K^{*0}$ and $\phi$ spin alignment measurements with respect to
the production plane in mid-central Au+Au collisions at
$\sqrt{s_{NN}}$ = 200~GeV together with the $\phi$ meson results
in p+p collisions at the same incident energy. As is the case for
our measurements with respect to the reaction plane, the
uncertainties in the measurement with respect to the production
plane are smaller for the $\phi$ than for the $K^{*0}$ meson, and
the statistical uncertainties are somewhat smaller than the
systematic uncertainty estimates. The $\phi$ point-to-point
systematic uncertainty estimate includes a dominant contribution
ranging from 0.030 to 0.045 due to residual background plus two
smaller contributions of 0.006--0.012 and 0.005--0.010 estimated
by varying the $K^{\pm}$ identification cut on $\langle dE/dx
\rangle$ from 2$\sigma$ to 1$\sigma$ and the fit range of the
$\phi$ meson invariant mass from 1.00--1.04~GeV/$c^{2}$ to
1.00--1.06~GeV/$c^{2}$. For the $K^{*0}$ we estimate a residual
background contribution to the point-to-point systematic
uncertainty ranging from 0.02 to 0.08 and about equal
contributions ranging from 0.01 to 0.08 by varying particle
identification criteria and analyzed rapidity. The Au+Au data for
$\rho_{00}$ are consistent with 1/3 to within 1--2 times the total
uncertainties, though the central values tend to increase with
decreasing $p_{T}$ for $p_{T}<2.0$\,GeV/c. The p+p results are
consistent with the Au+Au results and with 1/3. No conclusive
evidence is found for large polarization phenomena in the
production dynamics of vector mesons in the covered kinematic
region with the precision of current measurements. The p+p results
are in qualitative agreement with the suggested relation of vector
meson spin alignment with respect to the production plane and the
null results observed for the transverse spin asymmetries in
singly polarized p+p collisions at
mid-rapidity~\cite{STAR-ANMR,PHENIX-AN}. OPAL and DELPHI have
previously reported similar null results for the spin alignment of
the $K^ {*0}$ and $\phi$ mesons produced with small fractional
momenta ($x_p\leq0.3$, $x_p = p/p_{beam}$) in $e^ +e^-$
collisions~\cite{LEP-OPAL,LEP-DELPHI}, though the production and
fragmentation processes involved there are different from those at
RHIC.

\begin{table}
\centering \caption{The averaged spin density matrix elements
$\rho_{00}$ with respect to the production plane in mid-central
Au+Au collisions and the $\phi$ result in p+p collisions at
$\sqrt{s_{NN}}$ = 200 GeV.} \label{table2}
\begin{ruledtabular}
\begin{tabular}{c| c | c}
            &     $K^{*0}$    &    $\phi$
\\ \hline
$\rho_{00}(p_{T}<2.0$~GeV/c) & 0.43 $\pm$ 0.04 $\pm$ 0.09 & 0.42
$\pm$ 0.02 $\pm$ 0.04
\\ \hline
$\rho_{00}(p_{T}>2.0$~GeV/c) & 0.38 $\pm$ 0.04 $\pm$ 0.09 & 0.38
$\pm$ 0.03 $\pm$ 0.05
\\ \hline
$\rho_{00}(p_{T}<5.0$~GeV/c) & 0.42 $\pm$ 0.03 $\pm$ 0.09 & 0.41
$\pm$ 0.02 $\pm$ 0.04
\\ \hline
$\rho_{00}$(p+p) &  & 0.39 $\pm$ 0.03 $\pm$ 0.06
\end{tabular}
\end{ruledtabular}
\end{table}

In summary, we have presented the first measurements of spin
alignment for $K^{*0}$ and $\phi$ vector mesons at mid-rapidity at
RHIC. The results for the diagonal spin density matrix element
$\rho_{00}$ with respect to the reaction plane in Au+Au collisions
are found to be constant with $p_T$ in the measured region,
covering up to 5\,GeV/c, and constant with centrality. The data
are consistent with the unpolarized expectation of 1/3 and thus
provide no evidence for global spin alignment despite the large
orbital angular momentum in non-central Au+Au collisions at RHIC.
The results with respect to the production plane are found to be
less than two standard deviations above 1/3 in Au+Au collisions
and are consistent with the results in p+p collisions at the same
collisions energy. The measurements thus constrain the possible
size of polarization phenomena in the production dynamics of
vector mesons. Future measurements of polarization with respect to
the jet production plane are complementary to the current
measurements because they are not sensitive to the initial
conditions and may probe the system's mean free
path~\cite{Polarization-New}.

We thank the RHIC Operations Group and RCF at BNL, and the NERSC
Center at LBNL and the resources provided by the Open Science Grid
consortium for their support. This work was supported in part by
the Offices of NP and HEP within the U.S. DOE Office of Science,
the U.S. NSF, the Sloan Foundation, the BMBF of Germany,
CNRS/IN2P3, RA, RPL, and EMN of France, EPSRC of the United
Kingdom, FAPESP of Brazil, the Russian Ministry of Sci. and Tech.,
the NNSFC, CAS, MoST and MoE of China, IRP and GA of the Czech
Republic, FOM of the Netherlands, DAE, DST, and CSIR of the
Government of India, Swiss NSF, the Polish State Committee for
Scientific Research, Slovak Research and Development Agency, and
the Korea Sci. $\&$ Eng. Foundation.


\begin{thebibliography}{999}
\bibitem{LiangPRL} Z. Liang and X. Wang, {\it Phys. Rev. Lett.} {\bf 94}, 102301 (2005);
Erratum: {\it ibid.}, {\bf 96}, 039901 (2006).
\bibitem{Polarization-New} B. Betz et al., {\it Phys. Rev. C} {\bf 76}, 044901 (2007).
\bibitem{Polarization-New2} F. Becattini, F. Piccinini and J. Rizzo, {\it Phys. Rev. C} {\bf 77}, 024906 (2008).
\bibitem{LiangQM06} Z. Liang, J. Phys. G: Nucl. Part. Phys. {\bf 34}, S323 (2007).
\bibitem{Sergei} S. Voloshin, nucl-th/0410089.
\bibitem{LiangPLB} Z. Liang and X. Wang, {\it Phys. Lett.} {\bf B629}, 20 (2005).
\bibitem{Recom-1} R. Hwa et al., {\it Phys. Rev. C} {\bf 66}, 025205 (2002);
V. Greco et al., {\it Phys. Rev. Lett.} {\bf 90}, 202302 (2003);
R. Fries et al., {\it Phys. Rev. Lett.} {\bf 90}, 202303 (2003);
D. Molnar et al., {\it Phys. Rev. Lett.} {\bf 91}, 092301 (2003).
\bibitem{STAR-white-paper} J. Adams et al., {\it Nucl. Phys.} {\bf A757}, 102 (2005).
\bibitem{Hyperons-polarizations-list-theory} B. Andersson et al.,
{\it Phys. Lett.} {\bf B85}, 417 (1979); J. Szwed, {\it Phys.
Lett.} {\bf B105}, 403 (1981); L. Pondrom, {\it Phys. Rep.} {\bf
122}, 57 (1985); R. Barni et al., {\it Phys. Lett.} {\bf B296},
251 (1992); J. Soffer and N. Tornqvist, {\it Phys. Rev. Lett.}
{\bf 68}, 907 (1992).
\bibitem{Hyperons-polarizations-list-data} G. Bunce et al., {\it Phys. Rev. Lett.} {\bf 36}, 1113 (1976);
P.T. Cox et al., {\it Phys. Rev. Lett.} {\bf 46}, 877 (1981); R.
Rameika et al., {\it Phys. Rev. D} {\bf 33}, 3172 (1986); C.
Wilkinson et al., {\it Phys. Rev. Lett.} {\bf 58}, 855 (1987);
L.H. Trost et al., {\it Phys. Rev. D} {\bf 40}, 1703 (1989); J.
Duryea et al., {\it Phys. Rev. Lett.} {\bf 67}, 1193 (1991).
\bibitem{vector-meson-1} I. Ajinenko et al., {\it Z. Phys. C} {\bf 5}, 177 (1980);
M. Barth et al., {\it Nucl. Phys.} {\bf B223}, 296 (1983).
\bibitem{vector-meson-2} A.N. Aleev et al., {\it Phys. Lett.} {\bf B485}, 334 (2000).
\bibitem{LEP-DELPHI} P. Abreu et al., {\it Phys. Lett.} {\bf B406}, 271 (1997).
\bibitem{LEP-OPAL} K. Ackerstaff et al., {\it Phys. Lett.} {\bf B412}, 210 (1997);
{\it Z. Phys. C} {\bf 74}, 437 (1997).
\bibitem{QGP-Unpolarized} R. Stock et al., Proceeding of the Conference on Quark Matter Formation and Heavy Ion
Collisions, edited by M. Jacob and H. Satz, World Scientific
Singapore, 1982, p.557-582; A.D. Panagiotou, {\it Phys. Rev. C}
{\bf 33}, 1999 (1986); A. Ayala et al., {\it Phys. Rev. C} {\bf
65}, 024902 (2002); A. Ya. Berdnikov et al., {\it Acta Phys.
Hung.} {\bf A22}, 139 (2005).
\bibitem{Qinhua-vector-meson} Q. Xu and Z. Liang, {\it Phys. Rev.
D} {\bf 68}, 034023 (2003).
\bibitem{Fermilab-AN} A. Bravar et al., {\it Phys. Rev. Lett.} {\bf 77}, 2626 (1996).
\bibitem{STAR-AN} J. Adams et al., {\it Phys. Rev. Lett.} {\bf 92}, 171801 (2004).
\bibitem{PHENIX-AN} S.S. Adler et al., {\it Phys. Rev. Lett.} {\bf 95}, 202001 (2005).
\bibitem{STAR-ANMR} B.I. Abelev et al., {\it Phys. Rev. Lett.} {\bf 99}, 142003 (2007).
\bibitem{STAR-Lambda} B.I. Abelev et al., {\it Phys. Rev. C} {\bf 76}, 024915 (2007).
\bibitem{eqa-theta} K. Schilling el al., {\it Nucl. Phys. B} {\bf 15},
397 (1970); Erratum: {\it ibid.}, {\bf 18}, 332 (1970).
\bibitem{Art} A. Poskanzer and S. Voloshin, {\it Phys. Rev. C} {\bf 58}, 1671 (1998).
\bibitem{Primodial-production-1} Y.J. Pei, hep-ph/9703243; Y.J. Pei, {\it Z.Phys.C} {\bf 72}, 39 (1996).
\bibitem{Primodial-production-2} F. Becattini and U. W. Heinz, {\it Z. Phys. C} {\bf 76}, 269 (1997).
\bibitem{STAR-kstar} J. Adams et al., {\it Phys. Rev. C} {\bf 71}, 064902 (2005).
\bibitem{STAR-phi} J. Adams et al., {\it Phys. Lett.} {\bf B612}, 181 (2005).
\bibitem{STAR-TPC} M. Anderson et al., {\it Nucl. Instrum. Meth. A} {\bf 499}, 659 (2003).
\bibitem{dEdx-Bichsel} H. Bichsel, {\it Nucl. Instrum. Meth. A} {\bf 562}, 154 (2006).
\bibitem{STAR-v2} J. Adams et al., {\it Phys. Rev. Lett.} {\bf 95}, 122301 (2005), and references therein.
\bibitem{JGMa-Thesis} J. Ma, Ph.D. Thesis, University of California-Los Angeles (2006).
\bibitem{Haibin-Thesis} H. Zhang, Ph.D. Thesis, Yale University (2003).
\bibitem{Liang-Private} Z. Liang and X. Wang, private communication (2007).
\end{thebibliography}
\end{document}